\documentclass[12pt]{article} 
\usepackage[proportional,scaled=1.064]{erewhon}
\usepackage[T1]{fontenc}

\usepackage[top=1in, bottom=1in, left=1in, right=1in]{geometry} 
\usepackage[protrusion=true,expansion=true]{microtype} 
\usepackage{enumitem} 
\setenumerate{label=(\arabic*), wide=\parindent} 
\usepackage{setspace} 
\expandafter\def\expandafter\quote\expandafter{\quote\onehalfspacing} 
\usepackage{amsmath}
\usepackage[colorlinks=true, citecolor=blue!50!black, linkcolor=blue!50!black, urlcolor=blue!50!black]{hyperref}
\usepackage{xcolor}
\usepackage{todonotes}
\usepackage{fancyhdr} 
\usepackage{titlesec} 
\titlespacing*{\subsection}{\parindent}{.25in}{\wordsep}
\usepackage{amssymb, amsmath, mathrsfs} 
\usepackage{stmaryrd} 

\newcommand{\dd}{\mathrm d}
 \newcommand{\ii}{\mathrm i}
 \newcommand{\e}{\mathrm e}
 \newcommand{\Tr}{\mathrm Tr}
 \newcommand{\del}{\partial}

\begin{document}
\title{Missing the point in noncommutative geometry}
\author{Nick Huggett\textsuperscript{a}\footnote{huggett@uic.edu}~, Fedele Lizzi\textsuperscript{b,c,d}\footnote{fedele.lizzi@na.infn.it}~, and Tushar Menon\textsuperscript{e}\footnote{tv288@cam.ac.uk}}
\maketitle

\small
\begin{center}
    \textsuperscript{a} Department of Philosophy, University of Illinois at Chicago \\
    Chicago, IL, United States \\
    \bigskip{}
    \textsuperscript{b} Dipartimento di Fisica ``Ettore Pancini'', Universit\`a di Napoli {\sl Federico ~II}, Napoli, Italy \\
    \bigskip{}
  \textsuperscript{c} INFN, Sezione di Napoli, Italy \\
  \bigskip{}
    \textsuperscript{d} Departament de F\'{\i}sica Qu\`antica i Astrof\`{\i}sica and Institut de C\`{\i}ences del Cosmos (ICCUB), Universitat de Barcelona, Barcelona, Spain \\
    \bigskip{}
  \textsuperscript{e} Faculty of Philosophy, University of Cambridge, Cambridge, United Kingdom \\
    
\end{center} 
\date{}
\maketitle

\begin{abstract}
    Noncommutative geometries generalize standard smooth geometries, parametrizing the noncommutativity of dimensions with a fundamental quantity with the dimensions of area. The question arises then of whether the concept of a region smaller than the scale - and ultimately the concept of a point - makes sense in such a theory. We argue that it does not, in two interrelated ways. In the context of Connes' spectral triple approach, we show that arbitrarily small regions are not definable in the formal sense. While in the scalar field Moyal-Weyl approach, we show that they cannot be given an operational definition. We conclude that points do not exist in such geometries. We therefore investigate (a) the metaphysics of such a geometry, and (b) how the appearance of smooth manifold might be recovered as an approximation to a fundamental noncommutative geometry.
\end{abstract}

\maketitle
\tableofcontents

\section{Introduction}\label{intro}

In the vast landscape of contemporary theoretical physics, few research programmes come as close to engaging, as a matter of course, with traditional metaphysics as quantum gravity research programmes do. The two theoretical edifices that these programmes aim to unify (or replace)---quantum theory and relativistic gravitation theory---have been notoriously uncooperative with attempts at unification. It is not clear precisely which aspect of these theories is to blame, and what aspects ought to be held onto in future theories. Consequently, we are led back to traditional questions in the metaphysics of space and time. Questions like: what is the nature of space(time)? Are space(time) points fundamental? Is space(time) discrete? 

In this paper, we discuss a particular approach to the metaphysics of discrete space suggested by one popular family of approaches to quantum geometry that go under the name of \emph{noncommutative geometries}. From a philosophical perspective, attention to noncommutative field theories is valuable, because these theories allow us to embed our extant, well-confirmed physical theories in a broader logical landscape. Doing so allows us to unearth a number of tacit assumptions in our current physical theories that might otherwise have been invisible, or appeared as matters of necessity.

Our goal, therefore, is to introduce noncommutative geometry to a wider philosophical audience, by discussing three metaphysical puzzles about the nature of space and, in particular, indeterminacy of location to which these geometries give rise. We understand `indeterminacy of location' as referring to situations in which, for whatever reason, nature does not ascribe to a body a determinate a matter of fact about its spatial location below a particular scale. The first puzzle, accordingly, is to characterise this particular brand of metaphysical indeterminacy. This leads to the second puzzle of how one ought to think about the ontology of a theory that is based on a noncommutative geometry. The final puzzle is to account for our experience of spacetime as, at least approximately, being described by a commutative geometry.

There is a family of approaches to modelling indeterminacy in quantum mechanics mentioned according to which, if quantum mechanics is true, then particular facts about the world are `unsettled'---we can pose questions about the values of certain properties of systems such as, say, the $x$ and the $y$ components of spin, but nature itself does not determine the answers to such questions. Here, we focus on the subspecies of these approaches dubbed `supervaluationist' (see e.g. \cite{darby2010quantum, darby2019modelling}).\footnote{Since the `determinables-based' approaches (primarily associated with Wilson and collaborators, \cite{wilson2013determinable, calosi2019quantum}) also presuppose a topological manifold structure on the domain of discourse, the criticism we offer in this paper also targets the determinables-based approach. However, for dialectical clarity, we choose to focus only on the supervaluationist approaches.} We refer to this approach as modelling \emph{indeterminacy as underdetermination}: one considers various precisifications of the models of the physical system, and then models indeterminacy as the underdetermination of which of the precisifications truly represents the world.

Some sorts of quantum indeterminacy can plausibly be modelled as underdetermination, because ordinary quantum mechanics presupposes a continuous manifold of spatial points structured by some geometric relations (in non-relativistic quantum mechanics, these are the relations that constitute Galilean spacetime). Each precisification is itself antecedently meaningful, on the basis of a localisability thesis that we defend below. Noncommutative geometry, on our interpretation, does not have the resources to make meaningful claims about localisability below a certain magnitude. We therefore argue that the indeterminacy that results from a noncommutative approach to spatial geometry, as suggested by noncommutative geometric approaches to quantum geometry is of a different kind from what the supervaluationists consider. We call this \emph{indeterminacy as meaninglessness}. In this paper, we cash out `meaninglessness' in two distinct ways, depending on the approach to NCG: (i)\S \ref{spec} presents Alain Connes' spectral triple generalisation of Riemannian geometry, and characterises meaninglessness as undefinability; (ii)\S \ref{operation} presents a concrete representation of quantum theories in noncommutative space, and characterises meaningless as non-operationalisability. We then invoke an Occamist norm to link these semantic claims to our preferred metaphysical picture on which we deny the existence of spacetime points.\footnote{We do not intend to suggest that this is the only way to make sense of the pointlessness claim---for an alternative picture, see e.g. \cite{reyes}.}

Having established our argument for a fundamental metaphysics that eschews the concept of arbitrary localisability, we discuss some alternative views of the ontology of a noncommutative field theory in \S\ref{ontology}. One thing all of these proposals agree on is that the elements of the relevant noncommutative algebra should be treated as fundamental. The picture of a field as an ascription of properties to points (supplemented with some story, kinematical or dynamical, about how those points are related to each other) is untenable. While fields-first proposals have been in the philosophical literature for decades (Earman discussed so-called Leibniz algebras at least as far back as 1977 \cite{earman1977leibnizian}), they have been presented as \emph{alternatives} to standard ontologies for commutative theories like general relativity. In noncommutative field theories, fields-first interpretations are the only game in town.

In \S \ref{recovering}, we go on to examine a proposal for the recovery, from a noncommutative underlying geometry, of physical spacetime that is at least approximately commutative. In particular, we discuss a proposal that relies on structural features of quantum field theories to allow us, at least in some restricted but nonetheless physically salient circumstances, to recover a geometry that is approximately Minkowskian. The mismatch between the manifest and scientific images of space is an especially acute problem in the case of a theory with a putatively non-spatiotemporal fundamental ontology. This is because all evidence for a theory is ultimately given in terms of spatiotemporal data. In recovering the manifest image from NCG, at least in a restricted context, we counter the possibility that NCG is \emph{empirically incoherent}, to use a term introduced by Barrett \cite{barrett1999quantum} to describe theories whose truth undermines our justification for believing in their truth.

\section{Spectral triples}\label{spec}

A standard move in contemporary philosophy of spacetime is to model a spacetime theory as consisting of a smooth (i.e. infinitely differentiable), second countable, Hausdorff manifold on which are defined some tensor fields which encode spatial and temporal relations (in relativity theory, a Lorentzian metric tensor field) and some other tensor (or spinorial tensor) fields representing a matter distribution. The elements of the smooth manifold are typically treated as constituting the domain of discourse, call it $M$, of the theory; these elements, (commonly referred to as the `spacetime points'), are considered to be part of the fundamental, non-derived, non-emergent ontology of the theory. 

We take such a structure as the starting point for our discussion, using it to define a notion of localisability in \S\ref{mean}. However, as we show in \S\ref{spectralapproach}, such localisability is undefinable below a certain distance in the noncommutative space proposed by Alain Connes. In this sense, such distances are unphysical, and with them point regions.

\subsection{Meaning and definition}\label{mean}

In this paper we are concerned with physical, hence \emph{metrical} space; our conception of localisation is correspondingly metrical: localisation within some region of \emph{determinate size}. Thus our arguments will be to the effect that nothing can be localised in regions smaller than a certain size in our quantum spaces: that there are no smaller regions. (Insofar as our interest is in the status of points, which are closed intervals, our definition of localisation invokes boundaries of objects.)
  Of course metrical notions of localisation are familiar from the mereology literature: for example,`exact location', according to which `entity $x$ is exactly located at a region $y$ if and only if it has the same \emph{shape and size} as $y$'  (\cite{sep-location}. Metrical localisation is to be contrasted with thinner topological conceptions, say of proper or improper set inclusion. But since the spaces, classical and quantum, which we consider are metrical -- because they purport to represent physical space -- a metrical notion is appropriate in this context.

Therefore, in one dimension, an entity is exactly localised within some finite interval of length, call it $\delta$ (from which one can straightforwardly define the more useful notion of localisation to a finite area) iff the coordinate functions, $x^i$ associated with the boundaries of that object satisfy the following constraint:
\begin{equation}\label{1}
    \delta=|x^i(p)-x^i(q)|\geq 0
\end{equation}
where $|\cdot|$ is a given norm in $\mathbb{R}^n$, $\delta < \infty$ and $p,q\in M$ are its boundary points.Call the claim that it is possible to localise a body to an arbitrarily small interval the \emph{localisability thesis}.

The domain has metric structure, which is to say, enough structure to allow us to define a distance between any two of its elements. We begin by looking at how the story about this structure might be told in a mathematical textbook. We can define a property like location in the manner above because a coordinate function can be stipulated to be an isometry from the domain, $M$, to $\mathbb{R}$. If, in addition, we do not wish to privilege a particular position, and instead care only about distances between pairs of points (e.g. boundary points of objects), we need to associate with $M$ an entire equivalence class of coordinate functions, each of which agrees on the length between end points. 

In this way, our preferred coordinatisations can be thought of as inducing metric structure on $M$, inherited from the primitive metric structure of $\mathbb{R}$. And something similar is true of topological, smooth, linear or any other form of geometric or algebraic structure---more exotic structures can be imposed on $M$ by choosing different mathematical spaces as target of coordinate functions. Let us call such a space a \emph{structured domain}, and denote it as $M_s$. This induced metric structure on the structured domain allows us talk of separations between elements of $M_s$ in terms of separations of the images in $\mathbb{R}$ of those elements under the coordinate functions.

On this set up, the space of coordinate functions encodes certain facts about some of the structure that our theory recognises. For example, the fact that we associated $M$ with an equivalence class of coordinate functions, each element of which disagrees over the precise coordinate value to which a particular element of $M$ is mapped, means that we cannot identify absolute positions in $M$.\footnote{We should flag that we are treading in the vicinity of a debate over so-called `symmetry-to-reality inferences' (SRI) [see e.g. \cite{belot2003notes, dasgupta2013symmetry,ismael2003symmetry}]. This debate focusses on question of how we can use symmetries of a theory as a guide to its ontology, and in particular what, if anything, justifies an eliminativist metaphysical stance with regard to symmetry-variant structure. The concern is a potential circularity in the argument that symmetry-variant structure is unreal, and that real structure is symmetry-invariant. This is a deep and fraught question, and one on which we do not wish to take a stance in this paper. What we do argue for is primarily a semantic claim---the relation of arbitrarily small spatial separation is not definable in a theory whose space of privileged coordinate functions have a particular structure. We avoid circularity concerns by stipulating the structure on the space of coordinate functions, and examining the semantic and metaphysical consequences of this stipulation. We do not, for example, argue that the algebra of coordinates we focus on is the correct algebra \emph{because} it renders undefinable arbitrarily short spatial separations.} To put this another way, the labels of absolute positions are not invariant under the automorphisms of the space of privileged coordinate functions. Here, automorphisms of the structured domain should be understood as bijections from the space of coordinates to itself that preserve the space of privileged coordinate functions.

Adapting terminology from model-theory, we refer to this condition as \emph{undefinability} (cf. \cite{enderton2001mathematical}), and where definability is characterised thus:

\begin{description}
\item 
\noindent
A piece of mathematical structure is definable (relative to a structured domain) if and only if it is invariant under the automorphisms of that structured domain, 
\end{description}

\noindent  As is demonstrated in the example below, for a putatively spatial or spatiotemporal theory (i.e. a theory in which there are no `internal' degrees of freedom) we can always characterise a structured domain by restricting the class of privileged coordinate systems.

Consider a structured domain $M_s$ of uncountable cardinality whose automorphisms characterise it as a topological manifold (i.e. whose coordinate functions are stipulated to be homeomorphisms into $\mathbb{R}$). Let us structure this domain further by defining a set of maps, $g:N\rightarrow\mathbb{Z}$ from a countable proper subset of $M_s$ to the integers, $\mathbb{Z}$. Let us stipulate that these coordinate functions are isometries, thus imposing on $N$ a metric structure. Denote the smallest distance between any two points in $N$ as $\mu$. 

Now consider a topology-preserving map from $M$ to itself (i.e. a homeomorphism) which, in addition, preserves the distances between the elements of $N$ as determined by the discrete coordinate function $g$. Our structured domain $M_s$, is now characterised in terms of this more restricted class of coordinate systems. 

Consider the following set of pairs of elements $p,q\in M$, call it $R_{\mu}$:$\{\langle p,q\rangle | d(p,q)<\mu\}$. One characteristic feature of this set is that if one element of a pair that constitutes $R_\mu$ is contained in $N$, then the other is not. So this set is not invariant under all automorphisms of $M_s$: for example, a homeomorphism on $M$ which is the identity on $N$ unchanged, but permutes every other element of $M$. In this case, $R_{\mu}$ is not invariant under the automorphisms of $M_s$, so is not definable relative to $M_s$. Here, we simply cannot model indeterminacy of location as mere underdetermination between models each of which specifies an arbitrarily precise location, given our understanding of exact localisation, because we cannot associate a real number $\delta < \mu$ in accordance with equation \eqref{1}.

Our restriction to a domain with this privileged discrete structure was contrived. But it was contrived in order to make an important point about representation and indeterminacy; the examples we discuss in the rest of the paper will be more physically motivated. Quantum indeterminacy in position of a particle on the de Broglie-Bohm interpretation, for example, can be unproblematically represented by an underdetermination between different models, where each model has associated with it a perfectly determinate notion of location, derived from the definability of arbitrarily short lengths (more precisely, a family of two-place relations on $M$ corresponding to arbitrarily small separations). And this is perfectly permissible (indeed, encouraged) in a formalism which treats the structured domain  as a metric space. In what follows, however, we will argue that a quantum mechanical approach to geometry mandates a noncommutative, discrete structure on the domain of discourse. As a result, separations below certain scales (again, more precisely, a family of 2-place relations on $M$ corresponding to separations below a certain scale) will not be definable, so indeterminacy in position cannot be modelled by underdetermination between models with determinate notions of location. 

\subsection{The spectral approach}\label{spectralapproach}

It appears, therefore, that we have a problem. It is all very well to say that a metric space can be characterised by assigning a real number to each pair of its elements, but how does one do this systematically for a domain $M$ with uncountably many objects? We cannot simply list numbers associated with all pairs of elements in $M$. The standard move is to define a \emph{line element} on $M$, $ds^2=g_{ij}dx^i dx^j$, which represents, roughly, an infinitesimal displacement, and then integrate this line element along an arc that connects any two points in $M$ in order to assign to that pair some real number. This `arc-connectedness' is the basis of all differential geometry, and is an extremely powerful piece of mathematical technology. 

Unfortunately, this move brings with it a problem: this definition of length requires that we can define arbitrarily short lengths, i.e. that the space is arc-connected. This won't do--recall from the previous section that we had good reasons to believe that lengths \emph{above} a certain minimal scale ought to be definable even if lengths were not definable below that scale. In other words, our desideratum is the ability to define \emph{some} lengths in a space that is not arc-connected, so our definition of length cannot require that a space be arc-connected. How ought one proceed in light of this demand? 

Alain Connes' formalism of spectral triples promises to solve this problem. In order to assess the plausibility of this claim, we need to understand both the motivations and the mechanics behind this proposal. We begin by making precise the questions of interest to which the spectral triple formalism provides the answer:

\begin{description}
\item \textbf{Question 1:} What is the minimal structure of the domain of discourse required to make sense of locations?
\end{description}

As discussed earlier, in physical space a natural answer to this question is `the structure of a metric space, $\langle M, d\rangle$', where $d$ is the geodesic distance between any two elements in the domain of discourse. Differential geometry demonstrates that arc-connectedness (together with a few other assumptions) is sufficient to characterise a metric space.
Our aim in this section is to explicate a generalisable alternative characterisation, thus demonstrating that arc-connectedness is not necessary. In other words, we seek to answer in the affirmative the following question:

\begin{description}
\item \textbf{Question 2:} Can we turn a domain of discourse of uncountable cardinality into a metric space even if it is not assumed to be arc-connected?

\end{description}

The goal, then, is to recover a metric space $\langle M,d \rangle$ algebraically, from a starting point that does not assume any metric or topological structure on the structured domain. Connes \cite{connes1994noncommutative} proposes using \emph{spectral triples}, $\langle \mathscr{A}, \mathscr{H}, D\rangle$, where $\mathscr{A}$ is a particular kind of algebra over a field, known as a \emph{$C^{\star}$-algebra}\footnote{A $C^\star$-algebra is an algebra over a complex field, equipped with an \emph{inner involution}, denoted by $\star$, which generalises the operation of taking the complex conjugate of a number, and a \emph{norm}, which generalizes the modulus of the complex numbers, with respect to which the algebra is complete. Complex number themselves are the simplest example of $C^{\star}$-algebra.}, $\mathscr{H}$ is a Hilbert space, and $D$ is a particular self-adjoint operator over that Hilbert space, known as a \emph{Dirac operator}. The idea is simple: define, in algebraic terms, the structures that we know and love from differential geometry as special cases of these triples, check that we can still do differential geometry after this step, and then modify and deform these structures in such a way as to generalise differential geometry to unfamiliar domains. In particular, this will allow us to ascribe to a structured domain enough structure to define all and only lengths above a certain scale, satisfying our earlier desideratum.

We split the task of recasting differential geometry in the language of spectral triples into two steps:

\begin{description}
\item \textbf{Step 1:} Recover a topological manifold, $M$, from a spectral triple.
\item \textbf{Step 2:} Recover a geodesic distance function, $d$ on $M$, from a spectral triple.
\end{description}

Step 1 is completely straightforward and it builds on mathematical work in the forties, mainly a well-known representation theorem due to Gelfand and Naimark \cite{gelfand1943rings}, which we describe briefly. Consider a Hausdorff topological space: for simplicity we may think of a manifold, $M$, for instance the (2-dimensional) plane or sphere. Defined on it are the scalar fields, continuous functions that assign a complex number to each point: the set of such functions\footnote{For our purposes we need not go into detail as for the class of functions, continuous (for topology) or smooth (for differentiability) or other classes.} is known as $C(M)$. To understand the following, it is important to distinguish between such fields, which are functions over all space, from their values at a point: the former are complete `configurations' of individual point-values.\footnote{The usual notation for fields, e.g. $\Psi(x)$, tempts a conflation here, as the argument could be read as a particular value; but it instead indicates that we have a function over (co-ordinatized) points. Since that is understood, we might leave out the argument, and just denote the field (the function, the configuration) as $\Psi$. With respect to the set $C^\infty$, the fields are its elements, and their identities depend on the point-values (two fields are the same field, iff all their point values agree). }

Two scalar fields, $\phi$ and $\psi\in C(M)$, can be multiplied together in an obvious way to obtain a third, $\chi\in C(M)$ -- the value of $\chi$ at any point $p$, is just the ordinary product of the values of $\phi$ and $\psi$ at $p$: $\chi(p) = \phi(p)\cdot\psi(p)$. Such `pointwise' multiplication of fields is in fact so obvious as to almost be invisible: how could there be an alternative? In \S \ref{moweyl}, we shall see that there are alternative rules for multiplying fields, and indeed, they may even be more physical than pointwise multiplication.

Because ordinary multiplication is commutative, $a\cdot b=b\cdot a$, so is pointwise multiplication for elements of $C(M)$: $\phi\cdot\psi=\psi\cdot\phi$.The algebra contains a great deal of information about the space on which the fields live. In fact, the algebra contains \emph{all} the information that we typically take to characterize a topological space. Topology, understood as characterizing relations between points, can be reconstructed from purely algebraic data as maximal ideals\footnote{An ideal is a subalgebra such that the product of one of its elements by any element of the algebra still belongs to the ideal.}, the neighborood of a point can be likewise inferred from the relations among ideals. Global characteristics are also encoded in the algebra; for example a closed compact space (such as a circle) is described by an algebra which contains a multiplicative identity element. By contrast open spaces such as the real line are described by algebras which lack such an element. In short, this representation theorem states the logical equivalence of a space topology and its algebra $C(M)$.  

It's worth emphasizing the strength of this point, by reflecting on what is meant by an `algebra': nothing but a pattern of relations -- a structure -- with respect to some abstract operations. One might, for instance fully characterize an algebra by saying that there are two elements, $\{a,b\}$, and an operation $\circ$, such that $a\circ a=b\circ b=b$ and $a\circ b=b\circ a=a$ (and specifying that the operation is associative). What the elements are is not relevant, neither is the meaning of $\circ$; all that matters is how many elements, and what function on pairs of elements $\circ$ is. Of course, an algebra can have different concrete \emph{representations}: concretely, $a$ might be represented by the set of true propositions, and $b$ by the set of false propositions, in which case $\circ$ is represented by the boolean not-biconditional connective. But there are other representations: addition mod-2 for instance (and perhaps $a$ could be represented by the presence of a 30kg hemisphere of uranium 235, $b$ by the absence, and $\circ$ by the operation of putting together -- the critical mass of U$^{235}$ is 52kg!). These are different representations of a \emph{single} algebra, which captures their common structure.

It is not relevant that the concrete elements of the algebra are fields over the manifold, all that need be specified are their relations with respect to a binary operation. However, the scalar fields on a particular manifold define a specific algebra, and, according to the representation theorem, no other manifold has scalar fields with the same algebra. The point is that the algebra does all the work: there is nothing smuggled in about the manifold simply because we realize the algebra with fields over it. Additionally, every abstract commutative $C^\star$-algebra can be represented as an algebra of continuous functions $C(M)$ over some Hausdorff space $M$. So that settles our choice of algebra in the spectral triple: let $\mathscr{A}$ be $C(M)$, the algebra of continuous complex-valued functions over $M$. In general, $\mathscr{A}$ determines a Hilbert space (the $\mathscr{H}$ in the spectral triple) as well via a procedure known as the \emph{GNS construction}.\footnote{Briefly, this is achieved by treating the underlying vector space of $\mathscr{A}$ as its own representation space and using the representation, together with a specification of a complex-valued functional, $\omega$ (more on this object in \S\ref{operdefin}) to induce an inner product on the vector space. Finally, one takes care to ensure that the space is completed in the associated norm.} 

Step 2 is a little more involved, and it follows step 1 by nearly half a century, thanks to the work of Alain Connes and others. As we noted above, we cannot immediately recover, from the algebra of continuous functions, a metric manifold $\langle M, d\rangle$ in the way that we could recover a compact topological manifold $M$. One might wonder, however, if there is some subalgebra of $C(M)$ that  encodes metric facts about $M$. And there is, but this algebra can only be picked out if we allow ourselves the third piece of structure in a spectral triple, the \emph{Dirac operator}, $D$. If we stipulate that our domain of discourse, $\langle M, d\rangle$, has enough structure for us to define a notion of spacetime  spinors, roughly speaking, we can then define a differential operator, $D:=i\gamma^{\mu}\partial_{\mu}$ on the vector space of these spinors.\footnote{ Usually, in order to define spacetime spinors, one equips a differentiable manifold with a so-called `spin structure'. This is tantamount to defining a Lorentzian inner product on each tangent space. A spinor can then be thought of as an object that transforms in a particular representation---the \emph{spinor} representation---of the Lorentz group (for a rigorous definition, see e.g. \cite{wald1984general}). A well known consequence of this transformation behaviour is that spinor states are invariant under rotations of $4\pi$, but in general, not rotations of $2\pi$. The set of spinors forms a vector space, and it is on this vector space that the Dirac operator is defined. } With this extra structure, we have enough machinery to isolate a subalgebra of $\mathscr{A}$ that will allow us to recover $d$, the geodesic distance on $M$.

Consider the subalgebra of $C(M)$ known as the (algebra of) \emph{Lipschitz functions}, defined as follows:

\begin{description}
\item Given two metric spaces, $\langle M, d_M\rangle$ and the real line $\mathbb{R}$, the function $f:M\rightarrow \mathbb{R}$ is a \textbf{real-valued Lipschitz function} if and only if for all $x_1,x_2 \in M$ there exists a real-valued constant, $K$ such that:
\end{description}
\begin{equation}\label{lip}
    |f(x_1)-f(x_2)|\leq K d_M(x_1,x_2)
\end{equation}

Since Lipschitz functions can only be defined on metric spaces, the idea is that if we can find a subalgebra of $\mathscr{A}$ of Lipschitz functions, $\mathscr{A}_L$, we could use that algebra to reconstruct the geodesic distance on $M$. The problem thus splits into two parts:

\begin{description}
\item \textbf{Part 1}: Identify the subalgebra of $C(M)$ that constitutes the algebra of Lipschitz functions that we denote as $C_L(M)$.
\item \textbf{Part 2:} Identify the appropriate Lipschitz function such that for every pair of points on $M$ one can identify it with the geodesic distance between those two points using equation \eqref{lip}. 
\end{description}

For part 1, we start by considering a bounded measurable function, $a \in C(M)$, on $M$. From the GNS construction associated with this algebra, we know that it can be represented as an operator on a Hilbert space.\footnote{To cut down on unnecessarily complicated notation, we refer to this element as $a$, while remaining vigilant about correctly identifying which space these similarly labelled operators belong to.} There is a theorem that states that this function will be almost everywhere\footnote{i.e. everywhere except for a measurable set of Lebesgue measure zero.} equal to a Lipschitz function if and only if the commutator $[D,a]$ is bounded \cite{connes1994noncommutative}, where $D$ is the Dirac operator defined on the Hilbert space $\mathscr{H}$. So if we specify a Dirac operator, we can identify the algebra of Lipschitz functions, $C_L(M)$, a proper subalgebra of $C(M)$.

For part 2, we see directly from the definition of a Lipschitz function, that, for the value $K=1$, the supremum of the norm of the difference of image points  \emph{is} the geodesic distance. We are thus led to the following suggestion for the geodesic distance function on $M$:
\begin{equation}\label{dist}
    d(p,q)=\mathrm{sup}\{|a(p)-a(q)|; a \in C(M),\|[D,a]\|\leq 1\}
\end{equation}
where $|\cdot|$ is the $L^2$-norm on $\mathbb{C}$ and $\|\cdot\|$ is the norm on the Lipschitz algebra $C_L(M)$.
Note that we are defining a distance function on $M$ indirectly---by appealing to structure in the algebra $C_L(M)$. We need to establish that this, in fact, gives us the correct expression for the geodesic distance. The rigorous mathematical demonstration, detailed in \cite[Ch.~6]{connes1994noncommutative}, requires the introduction of more technical machinery than we have introduced here. The upshot is that, with the help of some mathematical footwork, \emph{when the algebra is commutative}, one can map each path between two points in the manifold to a norm in the Lipschitz algebra in such a way that the shortest path is mapped to the largest norm and the longest path the smallest norm. The geodesic distance in an arc connected space is then mapped to a supremum norm in the algebra. Now this link no longer exists when the algebra is noncommutative, but we can, nonetheless still speak of a geodesic distance expressed in terms of (only) the Lipschitz algebra norm.

The advantage of the use of spectral triples, in the context of the discussion in the previous section, is clear---we can, using equation (\eqref{dist}), define a notion of distance between two elements of a domain of discourse \emph{even when that domain is not assumed to be arc-connected}. We thus have the construction that we required in order to answer question 2 in the affirmative. Crucially, since we no longer need to assume $M$ is arc-connected, we can generalise the algebra, from the commutative $C(M)$ to a noncommutative $\mathscr{A}$. We do not need to worry that there is no longer a sensible notion of infinitesimal distance. All we need is a determinate specification of geodesic distance between elements of the subset of the domain of discourse \emph{for which the notion of separation is definable}. This specification does not need to piggyback on a specification of infinitesimal distance, and is thus still available to us in spaces characterised by having a noncommutative algebra of functions.

Denote the convex subsets of a generic algebra $\mathscr{A}$ as $S(\mathscr{A})$; and the `extreme boundary' of a convex set $K$, as $\partial_E (K)$.\footnote{A vector space $K$ is called a convex set iff $\forall v,w\in K$ and $t\in [0,1]$, $tv+(1-t)w\in K$. Its extreme boundary is the set of all $v\in K$ such that if $v=tw+(1-t)x$ for certain $w,x\in K$ and $t\in [0,1]$, then $v=w=x$.} Elements of the extreme boundary of $S(\mathscr{A})$ are also known as \emph{pure states}. It turns out that the space of pure states is also homeomorphic to $M$ when $\mathscr{A}$ is commutative. So now we have two spaces, $M$ and $\partial_E S(C(M))$, arrived at by different constructions, that are homeomorphic to each other. Let us focus on one important pair, $\partial_E S(C(M))$ and $M$. This isomorphism means that points in $M$ stand in a one-one correspondence with pure states. Consider the appropriate expression for $d$ on $\partial_E S(C(M))$:\footnote{It is worth noting that this expression gives us distances on the entire space of states on $S(C(M))$, not just the pure states.
However, in this section, we focus just on the subset of pure states.}

\begin{equation}\label{a}
    d_{\partial_E S(C(M))}(\alpha,\beta)=\mathrm{sup}\{|\alpha(a)-\beta(a)|; a \in \partial_E S(C(M)),\|[D,a]\|\leq 1\}
\end{equation}

All of this demonstrates that we should not be fooled into thinking that a geodesic distance can only be defined when $M$ and  $\partial_E S(\mathscr{A})$ are homeomorphic---once this isomorphism is broken by replacing a commutative algebra with a noncommutative algebra of functions on $M$, $\partial_E S(\mathscr{A})$ is still a metric space with a metric given by equation \eqref{dist}. But now, this metric space is no longer isometric to the space $\langle M, d_M\rangle$ (unsurprisingly, given that they are no longer even homeomorphic). Consequently, and crucially, the pure states are no longer identified with points in $M$ (the choice of terminology is not accidental---in quantum mechanics, these are the standard pure states that can be identified with rays in the system's Hilbert space). 

To reiterate, when $\mathscr{A}=C(M)$, which is commutative, the space of pure states, $\partial_E S(C(M))$ is homeomorphic (and can be made isometric) to the space $M$, so it does not matter which space we begin with. This is no longer true when $\mathscr{A}$ is noncommutative. In this case, the geodesic distance function on $\partial_E S(\mathscr{A})$ still maps pairs of pure states to real numbers, but the space of pure states, $\partial_E S(\mathscr{A})$ is no longer homeomorphic to the domain of the functions that constitute $\mathscr{A}$. The function define in equation \eqref{a} now identifies distances between pure states which cannot, in general, be interpreted as points of space.

To relate this construction to the discussion in \S \ref{mean}, let us consider the algebra of coordinate functions. The algebraic structure on this space privileges certain coordinate functions, and automorphisms of the algebra preserve this privileging, thus allowing us to structure the structured domain, $M_s$. For certain noncommutative algebras (for example, the noncommutative algebra of coordinate functions that we discuss in the second half of this paper), the set of pure states is not a topological manifold, \emph{a fortiori} cannot be interpreted as a set of points.  This is what underlies the `pointlessness' of NCG, as alluded to in several discussions of NCG, for example:

\begin{quote}
  The concept of a point becomes evanescent, and in some cases one is forced to abandon it altogether. \cite[p.~95]{lizzi2009noncommutative}
\end{quote}

 The conceptual shift in NCG is to treat the algebra $\mathscr{A}$ as fundamental, and the structure on the space $M$ as derived (we will explore some of the metaphysical consequences of this move in \S \ref{ontology}). In regimes where we need only focus on commuting algebras of observables, distances between pure states can, to whatever the appropriate degree of approximation, be identified with distances between locations.
 
We are now in a position to understand how this construction allows us to determine under what circumstances the 2-place `spatial separation' relation is definable. Define an automorphism, $h$ of a spectral triple as an automorphism of $\mathscr{A}$ such that leaves the Dirac operator invariant. To reiterate, the spectral triple is playing the same role---structuring the structured domain---as the privileged coordinate systems were in the simple example in \S \ref{mean}.

If $\mathscr{A}=C(M)$, then $D$ picks out the same distance function before and after the transformation. In other words, $h$ is an automorphism of $\langle C(M), \mathscr{H}, D\rangle$ iff
\begin{equation}
   d(p,q)=\mathrm{sup}\{|h(a(p))-h(a(q))|; a \in \partial_E S(C(M)),\|[h(D),h(a)]\|\leq 1\}
\end{equation}

\noindent where $d(p,q)$ is defined as in equation (\ref{dist}).

This $d(p,q)$ is equal to the standard Riemannian distance defined directly on $M$:

\begin{equation}\label{riem}
    d_M(p,q)=\mathrm{inf}\int_\gamma g_{ij}dx^i dx^j
\end{equation}

Consider a set of pairs of elements of $\langle M,d_M\rangle$:
\begin{equation}
R_\mu:= \{\langle p,q\rangle \in M|d_M(p,q)=\mu\} \  
\end{equation} 
\noindent where $d_M(p,q)=\mathrm{inf}\int_\gamma g_{ij}dx^i dx^j=\mu$ and $\mu\in \mathbb{R}$. This relation is invariant under an automorphism $h$ on $\langle M,d_M\rangle$. This automorphism will induce an automorphism $h_A$ on $\langle C(M),\mathscr{H},D\rangle$ such that:

\begin{equation}\mathrm{sup}\{|h_A(a(p))-h_A(a(q))|; a \in \partial_E S(C(M)),\|[h_A(D),h_A(a)]\|\leq 1\}=\mu.
\end{equation}

We can now define a relation $R_{\mu}^{C(M)}$ on $\partial_E S(C(M))$:

\begin{equation}\label{6}
    R_{\mu}^{C(M)}:=\{\langle \alpha,\beta \rangle \in \partial_E S(C(M))|  d_{\partial_E S(C(M))}(\alpha,\beta)=\mu\}
\end{equation}

This relation is invariant under all and only the automorphisms of $\partial_E S(C(M))$ induced by automorphisms that preserve $R_\mu$. So $R_{\mu}$ and $R_{\mu}^{C(M)}$ are equivalent, and it does not matter whether we use an algebraic or geometric description. They agree on the magnitudes that we are interested in here---lengths and areas, and consequently, as $\mu$ is made arbitrarily small, both $R_{\mu}^{C(M)}$ and $R_\mu$ can be thought of as picking out the same relation even though they are defined on different sets. We propose that a necessary condition for the equivalence of a relation on a normed space (like $\partial_E S(C(M))$) to a relation of spatial separation on a manifold $M$ is that there exist a homeomorphism between the two spaces (given our understanding of localisability, this fails to be a sufficient condition).

 Of course, the point of characterising the same structure in two different ways is that the new characterisation still applies when we leave the classical regime and instead consider noncommutative algebras of observables. Let $\theta$ quantify the magnitude of the noncommutativity of $\mathscr{A}$, i.e. $\forall \hat{x},\hat{y} \in \mathscr{A}, [\hat{x},\hat{y}]= i\theta$. Since $\theta$ is the commutator of distances, it is an area, which, for example, one could identify with the square of Planck's length $\sim 10^{-70}~m^2$.

 Let $\mathscr{A}$ be noncommutative, and at the same time, let us structure $M$ as a metric space, $\langle M,d_M\rangle$. In this case, as mentioned above, $M$ will no longer be isometric (or even homeomorphic) to the space of pure states, $ \partial_E S(\mathscr{A})$.  We can, of course, still define a relation $R_\mu$ on $\langle M,d_M\rangle$, because it remains invariant under automorphisms, $h$ of $\langle M,d_M\rangle$. But things change when we take the algebra $\mathscr{A}$ as fundamental. For commutative algebras, we could exploit the assumed homeomorphism between $\partial_E S(\mathscr{A})$ and $M$ to speak of two `equivalent' relations, one defined on $ \partial_E S(\mathscr{A})$, the other on $M$.  
 
Consider  relation $R^{\mathscr{A}}_{\mu}$, where $\mathscr{A}$ is noncommutative:
\begin{equation}\label{2}
R^{\mathscr{A}}_{\mu}:= \{\phi,\psi \in \partial_E S(\mathscr{A})|d_{\partial_E S(\mathscr{A})}(\phi,\psi)=\mu\} \  
\end{equation} 
 \noindent where $ d_{\partial_E S(\mathscr{A})}(\phi,\psi)$ is given by equation (\ref{dist}). 
 
While this relation is, indeed, invariant under automorphisms of the noncommutative spectral triple, $\langle \mathscr{A}, \mathscr{H}, D \rangle$, therefore definable, it is no longer equivalent to the relation $R_\mu$. This is because although the metric space of pure states $\langle \partial_E S(\mathscr{A}), d_{\partial_E S(\mathscr{A})}\rangle$ is invariant under spectral triple automorphisms, it is no longer homeomorphic to $\langle M,d_M\rangle$. Therefore it is no longer possible to assess whether the two relations are co-extensive: $R_\mu$ and $R^{\mathscr{A}}_{\mu}$ are incommensurable relations. This clashes with what we had identified as a necessary condition for $R^{\mathscr{A}}_{\mu}$ to represent a spatial separation of magnitude $\mu$: that $\partial_E S(\mathscr{A})$ is homeomorphic to $M$.

We can, however, restrict our attention either to commutative algebras, or to regimes in which the algebra of relevant observables can be treated as being commutative (i.e. the scale $\mu$ is much larger than the noncommutation factor $\theta$). In these scenarios, $R^{\mathscr{A}}_{\mu}$ can be seen to be equivalent to (or nearly equivalent to) $R_\mu$. We can therefore have localization within sufficiently large regions, but not within regions below a certain scale. The upshot of this discussion is that $R_\mu$, and hence our concept of localisation, is definable in a theory whose domain of discourse is a spectral triple only if the algebra is commutative. Since we can express all of the dynamically meaningful components of the noncommutative theory without making any reference to separations below the scale $\mu$, on Occamist grounds, we excise these putative regions -- including points -- of spacetime from our ontology.

\section{Operationalism}\label{operation}

Philosophers of physics with operationalist leanings might be sympathetic to but nonetheless unmoved by the argument in the previous section.  Operationalism is a view about how words that describe concepts acquire meaning, according to which `the concept is synonymous with the corresponding set of operations.'\cite{bridgman1927logic} From the perspective of a physical theory, rather than trying to imbue an abstract formalism with physical salience, it is built into the formalism from the beginning. 

Operationalism was introduced by Percy Bridgman, and most famously discussed in his \emph{The Logic of Modern Physics}. Although Bridgman himself does not present his view as such, it can be read as advancing a specific semantic claim about scientific words, namely that the meaning of a word is (in a sense to be made precise) completely exhausted by a specification of the operations that one would need to perform in order to measure a magnitude corresponding to that concept. Of particular relevance to the discussion in this paper is the following from Bridgman:

\begin{quote}
    If a specific question has meaning, it must be possible to find operations by which an answer may be given to it. It will be found in many cases that the operations cannot exist, and the question therefore has no meaning. \cite[p.~28]{bridgman1927logic}
\end{quote}

For various reasons, as a semantic theory, operationalism is no longer fashionable amongst contemporary philosophers. However, the empiricist spirit of operationalism still underwrites a standard approach among physicists and philosophers of physics for clarifying obscure concepts. This approach often proceeds via some form of thought experiment: think of Newton's globes as a way of understanding absolute space or the behaviour of light rays and test particles as a way of understanding the chronogeometric significance of the metric in general relativity.

It is in this spirit that, in this section, we present what might best be termed a `tempered operationalism', according to which a necessary condition on a concept having physical content is that it is possible, by the lights of physical theory, to describe a (perhaps idealised) measurement procedure for a magnitude associated with the concept. We refer to such concepts as \textbf{operationally definable}. To give operationalism substance, one has to specify what measuring operations are available; since we are interested in the possibility of operationalizing points of space, we will consider a (probabilistic) particle detector.

\subsection{The tempered operationalist approach to points}\label{temper}

Having clarified the sense in which we understand operationalism, in this section, we describe an idealised location measurement procedure. We argue that, on this setup, in ordinary classical field theory, localisations can be arbitrarily precise in principle, but trouble starts to brew in noncommutative geometries, leading to a violation of \textbf{operational definability}.

In brief, the argument as follows. In classical field theories on a commuting space, any uncertainty in measurement of location is down to technological or epistemic limitations. To account for these limitations, we associate what one might call an \emph{epistemic state} with a particle: a probability measure over some interval of space representing our uncertainty about the exact location---the \emph{ontic state}---of a particle. We then model an idealised measurement in such a way that the probability distribution that characterises our epistemic state becomes infinitely peaked at the particle location, in the limit that the uncertainty tends to zero. 

In theories of noncommutative space, we again assume---but this time for \emph{reductio}---that there is an ontic state corresponding to an arbitrarily precisely localised particle. We construct the analogue of an epistemic state: a density operator. We then attempt to localise this epistemic state to an arbitrarily small area and discover that this leads to ascriptions of negative probabilities. Since these measures are not elements of the state space, this signals a pathology. The only way to avoid this pathology, we argue, is to drop the assumption that there is an ontic state corresponding to an arbitrarily precisely located particle. Thus, even in principle, it is not possible, in a noncommutative space, to localise a particle below a certain area. Operationally, then, such areas---and \emph{a fortiori} \emph{points}---are undefinable.

\subsubsection{Classical space}\label{bd}

We will thus operationalize spatial regions and points in terms of particle location measurements; if measurements are only physically possible to finite precision, we argue that areas smaller than that scale do not exist. So first, consider a single particle prepared in a state localised to a region in $M$. In practice, there will be some uncertainty because of technological limitations in the preparation of the system. Second, we model a location measuring device: some apparatus which ``clicks'' if the particle is in a region, and is silent otherwise. Again, in practice there will be uncertainty in the measurement, because of our lack of full control of the dynamics. Because of both kinds of uncertainty---and for technical and dialectical reasons---we model both the state of the particle, and the measuring device with Gaussian functions. 

Consider, first, the commutative case. Here, the coordinate functions of the two dimensions commute and therefore decouple (we will shortly see why this is not the case in  noncommutative space), so it is sufficient to consider just one dimension. We represent the state of a particle as a \emph{normalised} Gaussian centered around a point $x_0$:
\begin{equation}
\rho=\frac{1}{\sqrt\pi\alpha}\ e^{-\frac{(x-x_0)^2}{\alpha^2}}
\end{equation}

\noindent where $\alpha$ is the width of the Gaussian. 

Observables are represented in this formalism as real functions of configuration space, say $f(x)$, and a measurement will give as average value
\newcommand{\vev}[1]{\left\langle#1\right\rangle}
$
\vev f_\rho= \int dx \,\rho(x) f(x).
$
The limit $\alpha\to0$ is well defined and
$
\lim_{\alpha\to0}\vev f_\rho=f(x_0) \label{pureclassicalstate}
$
For instance, for position, $x$:
\begin{equation}
\vev x_\rho=x_0\ ; \  \vev{x^2}_\rho=x_0^2+\frac{\alpha^2}2
\end{equation}
so that the uncertainty is
$
\Delta x=\frac\alpha{\sqrt 2}
$
The limit $\alpha\to 0$ is well defined, and in that limit, the state is perfectly localised in $x_0$.

We also model a position measuring device, with resolution $\beta$ as a Gaussian.
\begin{equation}
\sqrt2\cdot g_1(x)=e^{-\frac{(x-x_1)^2}{\beta^2}}
\end{equation}
The average value of the corresponding observable, for $\alpha=\beta$ is then

\begin{equation}
\label{eq:gaussvev}
\vev{g_1}_\rho= e^{-\frac{(x_0-x_1)^2}{2\alpha^2}}
\end{equation}

This quantity is very small unless $x_0\sim x_1$, i.e.\ the device is near the particle. When $|x-x_1|\ll\alpha$ this quantity is close to one, and goes to zero very fast as  $|x_0-x_1|$ grows. In other words, this observable indeed corresponds to a position measurement, capable of discriminating the location of a particle to arbitrary accuracy, as $\alpha\to0$. We will see later (\S\ref{moweyl}) that things are quite different in noncommutative geometry.

\subsubsection{Quantum kinematics}\label{quantumspace}

In the previous section, we were interested in the classical kinematics of location. Consequently, our states were just elements of some configuration space, observables were functions of these states, and measurement outcomes (magnitudes) were elements in the range of these functions. Quantum kinematics is different. The textbook story\footnote{as presented in e.g. \cite{prugovecki1982quantum}.} is that states are positive trace-class operators\footnote{A state represented by an operator in quantum mechanics is trace class if it has a well-defined (i.e. basis-independent) trace. If this trace is positive, then the operator is positive trace class.} on a Hilbert space, observables are self-adjoint operators on that Hilbert space and measurement outcomes are probabilities over magnitudes and are determined by the Born Rule.  In particular, the expectation value associated with some observable, call it $\hat{O}$ is:
\begin{equation}
    \langle \hat{O} \rangle_\rho= \mathrm{Tr}(\rho O)
\end{equation}
where $\rho$ is a state.

By assuming that $\rho$ is \emph{positive} trace class we require that it has no  negative eigenvalues, which would correspond to pathological (or contradictory) negative probabilities for physical measurements. Thus we designate the assumption the condition of state physicality:

\begin{description}
\item \textbf{State Physicality}: Physical states of a quantum system are (represented by) positive trace class operators on complex Hilbert spaces.
\end{description}

\noindent For our purpose the important consequence of this condition is that physical states cannot be represented by trace-class operators with negative eigenvalues. 

If we begin with a set of self-adjoint operators on a complex Hilbert space, we can define an operator norm on this set, and endow this set with the structure of a $C^\star-$algebra. We can incorporate some of the structure of the Born Rule into the definition of a state by defining states as maps from the $C^\star-$algebra of operators into the field of complex numbers. More precisely, given an algebra with a norm, a \emph{state} $\rho$ is a map from the elements of the algebra into the set of complex numbers with the following properties:
\begin{eqnarray*}
    \rho(\alpha_1 f_1+\alpha_2 f_2)&=&\alpha_1 \rho(f_1)+\alpha_2\rho(f_2)\nonumber\\
\rho(1)&=&1 \nonumber\\ 
\rho(f^+f)&\geq &0\nonumber\\ 
\|\rho\|\equiv\sup_{\|f\|\leq1}\rho(f^+f)&=&1 \label{stategeneric}
\end{eqnarray*}

$C^\star$-algebras allow us to express the kinematics of classical and quantum mechanics in broadly analogous ways. We should flag, however, the contentious nature of the appellation `quantum'. For some, e.g.~\cite{landsman2017foundations}, it is sufficient for a theory to be quantum that the algebra of observables be noncommutative. Others \cite{feintzeig2017theory}, argue that the quantum/classical divide is less clearly delineated, and lies on a continuous spectrum, where the actual divide is sensitive to other considerations. The cases that we consider in this section are noncommutative, but are not `quantum' in the narrow sense of imposing $\hbar$-dependent commutation relations on the canonical variables. In this paper, we remain neutral on the substantive question, but for clarity speak only of commutative and noncommutative field theories, reserving our use of `quantum' for later (\S\ref{finding}), for domains that can strictly be thought of as quantum mechanical. But the point here is that the framework just developed is appropriate for any situation with noncommutative observables, and so is  appropriate to the case of noncommutative \emph{space}, as we shall now see.

\subsection{Noncommutative space}\label{operdefin}

With the algebraic characterisation of geometry in mind, we now model the measurement procedure as we did in \S \ref{bd}, \emph{mutatis mutandis}, for a noncommutative space.\footnote{A discussion of the role of observers in a pointless space, of which the quantum phase space is an example, can be found in~\cite{Zalamea} and references therein.} We do not need the full structure of the spectral approach described in \S\ref{spectralapproach}, instead we work with a simple special case. We should stress at this stage that the lessons that we draw about measurements in this simplified model, based as they are on claims about the noncommutivity of the algebra of coordinates, generalise to more other systems modelled by noncommutative spectral triples.

Consider a simple toy example of a space in which what were previously coordinate functions (what we henceforth refer to as `base elements') are now self-adjoint operators with the commutator:
\begin{equation}
[\hat x,\hat y]=i \theta.
\end{equation}
As before, the area $\theta$ is the measure of the noncommutativity, analogous to $\hbar$, in the canonical commutation relations. The algebra of \emph{functions} of $\hat x$ and $\hat y$---the field algebra $\mathscr{A}$---which in the classical case was commutative, is now noncommutative. In particular, their polynomials are noncommutative elements of the algebra. 

So we seek a representation of the algebra, satisfying the quantum kinematics described above, via a `quantization map' that associates to any function an operator. There is some freedom in this choice, and usually one uses the \emph{Weyl map}, which associates self-adjoint operators to real functions. The definition is made via the Fourier transform
\begin{equation}
\tilde f(\xi,\eta)=\int \dd x \dd y f(x,y) \e^{\frac\ii\theta (\xi x+ \eta y)}
\end{equation}
the operator corresponding to the function $f(x,y)$ is therefore
\begin{equation}
\hat F=\int \dd \xi \dd\eta \tilde f(\xi,\eta) \e^{-\frac\ii\theta (\xi \hat x+ \eta \hat y)}.
\end{equation}
The inverse map, which associates functions to operators, is called the \emph{Wigner map}:
\begin{equation}
\tilde f(\xi,\eta)=\Tr F  \e^{\frac\ii\theta (\xi \hat x+ \eta \hat y)} \label{commrel}
\end{equation}

The operators we obtain can always be represented as acting on a separable Hilbert space, so both operators and states are represented as (infinite dimensional) \emph{matrices}.  Then the requirements~\eqref{stategeneric} translate to the claim that we can associate with a state an Hermitean matrix $\hat\rho$ with positive eigenvalues and {Tr}$\hat{\rho}=1$. This is standardly referred to as a density matrix.

Consider now a classical Gaussian epistemic state peaked around the origin and of width $\alpha$:
\begin{equation}
\rho=\frac1{\alpha^2\pi}e^{-\frac{x^2+y^2}{\alpha^2}} \label{gaussianstate}
\end{equation}

The quantization map associates an operator to this function, and we may write this operator in the $x$ basis as:\footnote{Remember that $x$ and $y$ do not commute, and therefore do not have a simultaneous basis, hence we cannot write the operator using both of them.}
\begin{equation}
\hat\rho=\frac1{\alpha^2\pi}\int \dd\eta \dd\xi \e^{-\frac{\xi^2+\theta\eta^2}{\alpha^2}}e^{-1(x\eta+\del_x\xi)}
\end{equation} 
This operator has well defined action on functions of $x$.

This matrix has been calculated (\cite{derezinski2017quantization,Leonetesi}, and found to be:

\begin{equation}
\hat\rho=\begin{pmatrix}
\frac{2\theta}{\alpha^2+\theta} & 0 & 0 & \cdots\\
0 & \frac{2\theta}{\alpha^2+\theta} \left(\frac{\alpha^2-\theta}{\alpha^2+\theta}\right) & 0 & \cdots\\
0& 0 &  \frac{2\theta}{\alpha^2+\theta} \left(\frac{\alpha^2-\theta}{\alpha^2+\theta}\right)^2 & \cdots\\
\vdots & \vdots & \vdots & \ddots 
\end{pmatrix}
\end{equation}

This operator is trace class, with Tr$\hat\rho=1$ in all cases. So it satisfies the first condition of state physicality. But under certain circumstances, it fails the second: for certain values of $\alpha$, it gives rise to negative expectation values for any observable. This can be inferred from the eigenvalues of this operator, for particular values of $\alpha$ and $\theta$.
\begin{itemize}

\item[$\alpha>\sqrt\theta$] The state is not particularly localised, and $\hat\rho$ is a density matrix, with all eigenvalues positive and smaller than 1.

\item[$\alpha=\sqrt\theta$] The state is a density matrix with the first eigenvalue equal to 1, and the rest vanishing.

\item[$\alpha<\sqrt\theta$] The matrix \emph{is not} a density matrix, the first eigenvalue is larger than 1, and negative eigenvalues appear, which lead to negative expectation values for all operators. Thus the operator violates \textbf{state physicality}.
\end{itemize}

Thus the quantum state corresponding to a classical epistemic state localised to an area below $\sqrt\theta$ does not correspond an ontic quantum state---the associated density operator does not represent a physical state.

To better understand this result, consider the position measurements that we take to operationalize position in the noncommutative framework. $\vev{\hat x^2} = 0$, which reveals no problem, but if we consider higher powers we find:

\begin{equation}
\vev{\hat x^4}=\frac3{16}(2\alpha^4-\alpha^2\theta-\theta^2)
\end{equation}
which---impossibly---is negative for $\alpha<\sqrt\theta$. Moreover, that the value of $\vev{\hat x}$ is positive is in fact an artifact of the symmetric state we chose; a different $\rho$, not symmetric around the origin, would have shown pathological behaviour even for a direct position measurement. In \S\ref{moweyl} we will introduce the formalism necessary to also see the noncommutative analogue of \eqref{eq:gaussvev}, a position measurement involving an imprecise device. We will see in more concrete terms why we cannot operationalize position measurements: attempts to measure below a certain scale are frustrated.

We conclude that measurements attempting to localize particles to a linear scale smaller than $\sqrt\theta$ are unphysical; of course, this result is a reflection of the fact that there are no ontic states localizing particles below this scale. Then, from our tempered operationalism, we conclude that space in fact has no regions smaller than this scale -- and in particular, is `pointless'.

\section{Ontology}\label{ontology}

We have argued that the notions of point, or region smaller than the commutation scale are undefinable in noncommutative geometry; undefinable in the formal sense of definition in a structured domain, and undefinable in the physical sense of tempered operationalism. All we have is a (metric) space of (pure) states, and these can only be identified as being pointlike if the algebra of fields is commutative. In that case, the traditional conception of the manifold fails too, and with it the `fields as properties of spatiotemporal points' view about the ontology of a field theory (as endorsed by, e.g. Hartry Field \cite{field1984can}). Thus the question of the interpretation of theories of noncommutative geonetry arises: what kind of world, what kind of basic elements, does it describe? One immediate thought is to develop a structure of spacetime by composing \emph{regions} each of which is of size above the commutation scale. The issue with this proposal is that the set theoretic closure of such open sets includes regions smaller than the commutation scale; so such an interpretation would require non-standard laws of composition. We take such issues as indicating that such an approach would shoehorn classical notions into what is, essentially, a quantum theory. So we will instead propose that the algebra, $\mathscr{A}$, whose elements are not intrinsically spatiotemporal at all, is itself the basic ontology. Naturally, this will require some elaboration!

Note that until now, we have only spoken of spatial, rather than spatiotemporal noncommutivity. From this point on, we also indulge in talk of noncommutative `spacetime'. But note that we always take the time coordinate to commute with the spatial coordinates (even when the latter do not commute with each other).\footnote{Noncommuting time is a delicate issue, technically and conceptually. See \cite{hugeal12,Bozkaya:2002at} for discussion.} We will also refer to noncommutative geometry as `NCG' for brevity.

In Book IV, Chapter 1 of his \textit{Physics} Aristotle offers the view that existence requires being somewhere: everything that is, has a place. (He is setting up the question, attributed to Zeno, of where places are, if they exist.) This idea is intuitive: the world seems fundamentally spatial, and it starts to capture the idea that `real' things can be interacted with, by traveling to them. If one accepts such a view of existence, then it becomes impossible to take the algebraic formulation of the theory as giving an ontology for the world, because it does away with space as a fundamental object. Modern philosophers are more likely to take a logical view of existence, perhaps adopting Quine's view that to be is just to be the value of bound variable in a true theory. At any rate, seems inadvisable to reject the algebraic formulation out of hand because of a view of existence. 

Related to the idea that ties existence to space, is the idea that comprehension requires the spatial. Maudlin \cite{maudlin2007metaphysics}, following Barrett \cite{barrett1999quantum}, expressed a concern that theories without fundamental spatiotemporal quantities could not be connected to experiments, which immediately concern local beables. But a vaguer objection that only spatiotemporal theories can be properly `understood' perhaps remains; useful predictions might be possible, but otherwise a theory can only be an instrument, not comprehendible by us. Kant had a view like this of course, which influenced Maxwell in the construction of electromagnetism; and such claims were made by Schr\"odinger in his arguments with Heisenberg. See (\cite{de2001spacetime}) for some history of the topic. But such instrumentalism is no more of a threat to an algebraic interpretation than Aristotle's spatial view of ontology. First, one can make the same kind of move to logic as before. Suppose a theory is algebraic rather than geometric, then it may not give an image that is easily visualized by the human mind, but it still can provide understanding in the sense of systematizing the connections between different parts of nature; between the quantum and gravitational realms, ideally. That is, our ability to understand formal systems that aren't spatial does plausibly give us the ability to understand physical systems that are not spatial. Second, there is a sense of `understanding' that indicates facility with a theory rather than any sense that its models can be visualised. For instance, de Regt  (\cite{de2001spacetime}) develops Feynman's view that understanding a formalism is a matter of `knowing' what the solution to a problem should be without having to compute it explicitly. But as de Regt points out, while our geometric intuitions are a fruitful resource for `seeing solutions', they need not be the only one; again, familiarity with an algebra also allows one to anticipate when algebraic relations hold without explicit calculation. 

So there do not seem to be in principle barriers to developing an algebraic theory. All the same, commutative space seemingly provides a very useful tool for investigating ontology, insofar as individuals can be separately localized, and the parts of space give a way of distinguishing parts of individuals. (Of course, even non-relativistic quantum non-locality makes this road to ontology treacherous at best.) Or again, \emph{physical} reality is often tied to causal connectedness, which in turn is most readily understood in terms of effects propagating in space. So despite our metaphysical sophistication it is still puzzling to know where to start in talking about the ontology of a theory like NCG, in which the familiar spatial handles are missing. What, then is there? How can we discern a coherent ontology from the theory?

Since we are considering a theory that replaces differential geometry with algebra it will be useful for the following to bear in mind two kinds of interpretational moves made regarding space. There are two main questions at stake: (i) to what extent, if any, are the points of a differential manifold real, physical objects, akin to material systems? (ii) What aspects of spatiotemporal structure, such as topological and geometrical relations, are fundamental (capable, for instance, of providing `deep' explanations)? Especially, following the (re-)introduction of the `hole argument' (\cite{earman1987price}), the locus of philosophical debate was on the first question: the `manifold substantivalist', who holds the points to be physically real, is faced with versions of Leibniz's shift arguments, in which one imagines the material content of the universe rearranged in spacetime. Earman and Norton's argument makes the point especially sharp in theories with dynamical geometries, such as GR, since then the problem of indeterminism can (arguably) be added to that of underdetermination: because of diffeomorphism symmetry there are solutions of the Einstein field equation that agree up to a given Cauchy surface, but differ by a diffeomorphism after. 

Recent work has largely focussed on the second question. The idea of `dynamical interpretations' of spacetime theories is that certain spatiotemporal structures (particularly affine and metrical ones) are not fundamental, but merely represent, say, the symmetries of the laws of material systems; hence `real' explanations (of time dilation, for instance) are in terms of how systems behave according to physical laws, not geometry (for instance, (\cite{brown2005space}); such interpretations are discussed in (\cite{huggett2009essay, readbrowndynamical}). With all this in mind, consider some possible interpretational strategies; we do not aim to adjudicate between them, but merely demonstrate the possibility of an algebraic interpretation.

First, in the formalism of NCG, instead of spatial points and their relations, we have elements of an algebra and their relations; this observation suggests that the elements could be thought of, metaphysically, along the lines of points. To pursue this idea more concretely let's take the algebra $\mathscr{A}$ to be $\mathcal{R}^d_\theta$, the algebra of polynomials of the noncommuting base elements, $\hat{x}$ and $\hat{y}$.\footnote{Standing in for physically realistic algebras of more complex objects like tensors and spinors.} `Algebraic substantivalism' then attributes to the elements of $\mathcal{R}^d_\theta$ the same kind of `physical reality' that manifold substantivalism attributes to points. To be a little more careful, just as the latter view takes mathematical points to represent, more-or-less literally, physical points, so algebraic substantivalism takes the elements of the mathematical algebra to represent, more-or-less literally, physical objects, which we shall continue to call `fields'. To be clear, the mathematical representation of the NCG is not itself something physical, but, according to substantivalism, what it represents is.

The idea that the points of a mathematical spacetime manifold could represent points of physical spacetime seems to be a natural one; at least philosophers (including Newton and Leibniz) have taken it (or something like it) to be a view worth defending or disputing. It appears that applying parallel reasoning in the parallel case of NCG feels less natural. However, the only difference between the two cases lies in the non-spatiality of the fields. But that is no reason to reify in one case and not the other: as far as existence goes, we have already rejected spatiality as a condition. And while non-spatiality makes the fields less immediately connected to objects of experience, we shall see below how NCG does connect with experience. In other words, however manifold substantivalism views points, algebraic substantivalism views elements of the algebra; understand one and you understand the other.

John Earman in \cite[\S9.9]{earman1989world}  proposed using such an interpretation to advance the substantivalist-antisubstantivalist debate, based an algebraic formulation of general relativity, known as an `Einstein algebra', due to Robert Geroch \cite{geroch72}. Because an Einstein algebra fixes a spacetime only up to diffeomorphism, it seems (but see below) that symmetric situations get the very same algebraic descriptions; so Earman suggested that an interpretation that takes the algebra as fundamental would thus avoid the hole argument. Moreover, Earman described the interpretation as substantival `at a deeper level'.

For an alternative interpretation, suppose one has a noncommutative field theory (NCFT) of a scalar field: suppose that the dynamical object of the theory is a scalar field, as in electromagnetism one has a theory of an antisymmetric tensor field. For want of a better term, call this a `material field', to distinguish it from the fields of the algebraic geometry (though it is only  `material' in the same sense that the electromagnetic field is). Then, each (algebraic) field corresponds to a \emph{possible} state of the material field, and can be interpreted to \emph{be} such. This move is reminiscent of the anti-substantivalist proposal that points are merely possible locations, not physical objects themselves; the typical response to this suggestion is that it simply introduces new entities with all the troubling features of points, and so the difference is too small to generate a truly distinct view in the interpretation of spacetime theories. But note that in scalar NCFT things are potentially more promising in that no new possibilia are proposed, because the material field is already assumed to be a determinable, with many possible determinate `configurations', with a noncommutative structure. Of course one might say that locations are also possible states of spacetime objects, but in that case there is the option of taking \emph{relative positions} to capture locations. There seems to be no corresponding move for the states of a noncommutative scalar field.\footnote{(\cite{dieks2001space}) proposes an approach along these lines for spacetime in ordinary QFT, in which points are identified with sets of local observables, themselves taken as possible properties of systems.}

A third option for interpretation is suggested by Bain (\cite{bain2003einstein}) in the  context of the Einstein algebra formulation of general relativity. In that case diffeomorphisms correspond to automorphisms of the algebra, and so Leibniz shifts have an algebraic counterpart: there are maps of the algebra onto itself that preserve the algebra, and hence the geometry. Do we not \cite{rynasiewicz1992rings} recapitulate hole indeterminacy at the algebraic level? Not, it seems, if we pose the question of determinism as follows: for any two algebraic models whose representations, restricted to $t<T$, are the same,  are they the same representations \emph{tout court}? Given the representation theorems about Einstein algebras, the answer is `yes'. 

However, Bain, is unconvinced by this response. He proposes that the fields only have identities in virtue of their algebraic relations to one another -- a version of `structuralism', since the fields become bare relata for the essential, algebraic, content of the theory. In this case, since the structure is preserved by automorphisms, such shifts make no difference; the analogy to similar moves in the hole argument -- or in response to other issues arising from shifts -- should be clear. One even (in principle) could go a step further, and treat the elements of the algebra as purely formal, not representing any physical `fields', even with weakened identity conditions. What would be left would be pure structure. It seems that all of these structuralist moves could be made just as well in the noncommutative as commutative cases (though the question of indeterminism remains).

Algebraic substantivalism and structuralism, and the idea that fields are states of a scalar field focus on the ontology of the elements of the algebra, as similar views concern the status of points in standard spacetime theories, but perhaps one could also attempt a `dynamical' interpretation of NCFT. Such a view would take the `material' noncommuting field as the fundamental thing, and view the algebra (certainly the algebraic relations, but potentially also its elements) as merely representing something about its equations of motion. How such a view might be developed, how plausible it might be, and how it might relate to the others described here, are questions that will have to be addressed at another time.

\section{Recovering space from NCG}\label{recovering}

All of the mathematical generalisations, and suggestive analogies might be of only abstract interest, if it weren't the case that the framework of modern \emph{physics}, also survives the transition to noncommutativity. The geometry and calculus involved can be given algebraic form, and survive. Importantly, the Lagrangian that characterizes any theory, can also be fully rendered in algebraic terms, even in the noncommutative case: so we have physics in a noncommutative geometry, specifically, a `noncommutative field theory' (though note, as in \S \ref{operation}, it is the geometry that the fields `inhabit' that is noncommutative, as the fields themselves have not been quantized at this stage). Moreover, other important pillars of modern physics, like Noether's Theorem, also survive (it only requires that the algebra be associative): hence the central importance of conserved currents remains. (And gauge fields also exist, though importantly the distinction between `internal' gauge symmetries, and `external' spatial symmetries is blurred.) In this section, we discuss a proposal for the recovery of the manifest image of a classical (commuting) space from a noncommutative field theory. We begin, in \S \ref{moweyl}, by introducing two formally equivalent representations of a noncommutative field theory. Then, in \S \ref{finding} we discuss and assess a proposal about the emergence of ordinary spacetime from this algebraically construed, fundamentally noncommutative field theory. We conclude in \S \ref{empiricalcoh} with a short reflection on the problem of empirical incoherence that we touched upon earlier.

\subsection{The Moyal and Weyl representations}\label{moweyl}

Let us try to understand how spacetime might emerge from a theory of non-spatial degrees of freedom. On the one hand, prima facie we have no degrees of freedom that are intuitively `spatial': no point-valued fields, but instead the state-values are simply elements of the algebra of functions multipled via the $\star$ product defined below.  $\mathcal{R}_{\theta}^d$. While on the other, the theory apparently contains enough structure that one successfully might connect it (in some kind of limit) with familiar physics, in a classical, phenomenal, spacetime. (Note that in this section, the fields, while noncommuting, are still not fully quantized: we do not have $\hbar$-featuring commutation relations on the canonical variables. We turn to a fully quantum formulation in \S\ref{finding}.)

Earlier (\S \ref{operdefin}), we gave an operator representation of the noncommuting algebra, but we can equivalently represent it in terms of functions on a manifold, by deforming pointwise mulitplication to a new product. For instance, in our toy case, with the algebra  $\mathcal{R}^d_\theta$, we take the usual smooth coordinate functions $x$, $y$, etc as base elements, and introduce the Gr\"onewold-Moyal---or simply Moyal---product:\footnote{See \cite{szabo2003quantum} for a full treatment. Here we continue to assume flat, infinite space, but the representation extends to more complex cases. We also restrict attention to fields that vanish smoothly at infinity (so that the `physicists fundamental theorem of the calculus' reads $\int\mathrm{d}x\ \mathrm{d}f(x)/\mathrm{d}x=0$). This restriction is a common but notable assumption in physics: on the one hand it is justified locally by the assumption that arbitrarily distant differences are irrelevant; on the other hand it raises questions about the universality of physical theories.}

\begin{equation}
\label{eq:moyal}
\phi(x)\star \psi(x) = \phi(x)\cdot\psi(x) + \sum^\infty_{n=1}(\frac{i}{2})^n\frac{1}{n!}\theta^{i_1j_1}\dots\theta^{i_nj_n}\partial_{i_1}\dots\partial_{i_n}\phi(x)\cdot\partial_{j_1}\dots\partial_{j_n}\psi(x).
\end{equation}
Clearly the $\star$-product contains new terms in addition to ordinary multiplication.\footnote{Note that this star operation has nothing to do with the involution operation on $\star-$algebras. The product is not unique, and there are several others which reproduce the commutation relation~\eqref{commrel} (all of the translation invariant one are described in~\cite{Galluccio:2008wk}.} One just needs to observe that the new terms form an (infinite) sum of derivatives with respect to the coordinates, weighted by the elements of $\theta$. For instance,
\begin{equation}
\label{eq:x*y}
\hat x\hat{y}\to x\star y = xy + \frac{i}{2}\theta^{xy} \neq xy.
\end{equation}
We can use (\ref{eq:x*y}) to verify the relevant commutator in the Weyl transform:
\begin{equation}
[x,y]_\star \equiv x\star y - y\star x = xy-yx+(\frac{i}{2})\theta^{xy} - (\frac{i}{2})\theta^{yx} = i\theta^{xy},
\end{equation}
by the antisymmetry of $\theta^{xy}$. Thus commuting coordinates do indeed have the algebra of the noncommutative theory, \emph{with respect to the Moyal product}. 

Another instructive example is the product of Gaussians, which we might take to represent (as before) a localized `particle' $\rho$ and measuring device $g$. Supposing both have width $\alpha$, and locations $x_1$ and $x_2$, respectively, the product is:
\begin{equation}\label{eq:noncomgauss}
 g\star\rho=
\frac{\alpha ^4 \exp
   \left(-\frac{\alpha ^2
   (x_1^2+x_2^2)-2 x_1 \left(\alpha ^2
   x+2 \ii \theta  y\right)-2x_2
   \left(
   \alpha ^2 x-2 \ii \theta 
   y\right)+2 \alpha ^2
   \left(x^2+y^2\right)}{\alpha ^4+4 \theta
   ^2}\right)}{ \left(\alpha ^4+4
   \theta ^2\right)}
\end{equation}
Note that the product is not commutative because of the presence of the phase $2 \ii\theta y (x_1-x_2)$ which is not symmetric  under the exchange $x_1\leftrightarrow x_2$, therefore $g\star\rho\neq\rho\star g$. Moreover, the result, which one whould like to interpret as a probability, is a complex number. This latter aspect could in principle be resolved taking the modulus, but it is clear that the ``classical'' interpretation is becoming untenable.

\begin{figure}[h]\center
\includegraphics[width=0.45 \textwidth]{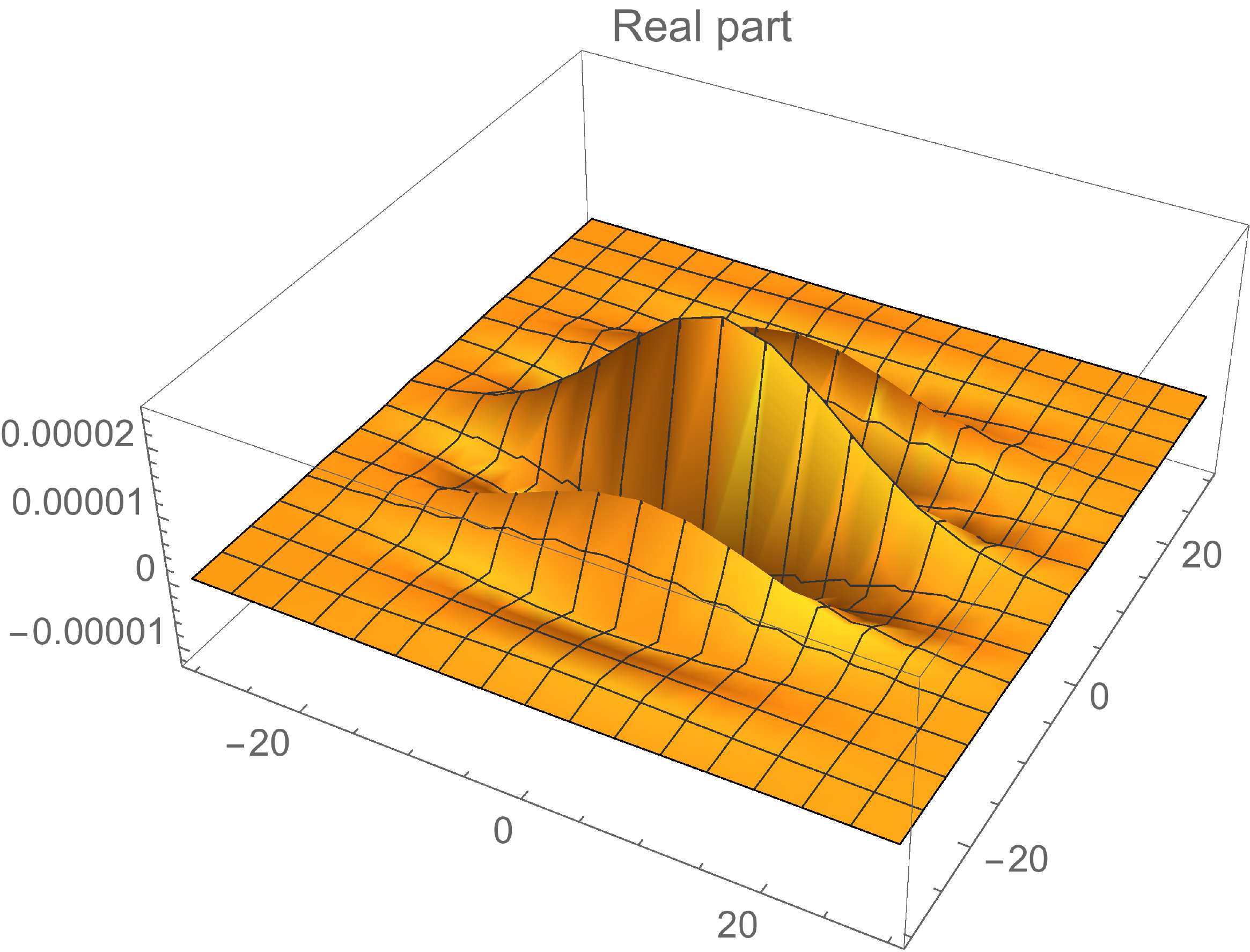}\qquad
\includegraphics[width=0.45 \textwidth]{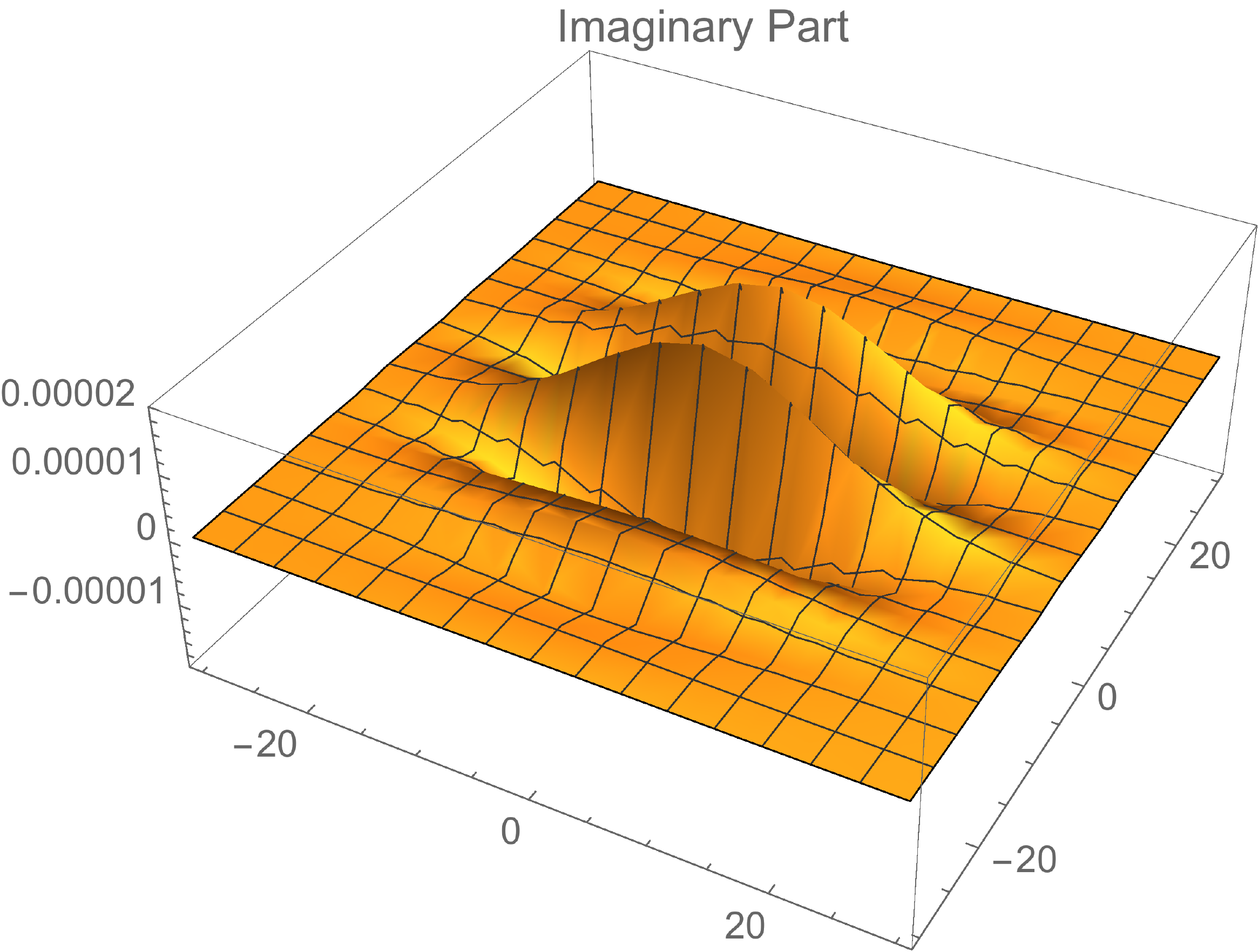}
\caption{\sl The real and imaginary parts of (\ref{eq:noncomgauss}) for width $0.1\sqrt\theta$,  at the distance $0.5\sqrt\theta$.}
\label{real2}
\end{figure}

We can also compare it with the product of classical Gaussians, to see the difference that noncommutativity makes. Let the Gaussians be sharply peaked, with $\alpha=0.1\sqrt\theta$), at the distance $0.5\sqrt\theta$. In the classical case the ordinary commutative product is practically zero everywhere (too small to be plotted), and the expectation value \ref{eq:gaussvev} is effectively zero. However, things are quite different for noncommuting fields, and the star product is nonvanishing, and quite spread, as can be seen in figure \ref{real2}. Although the integral which would gives rise to the expectation value in this case does not change,\footnote{This is a particularity of the Moyal product as opposed to other noncommutative geometries.}, all of the other moments of the probability change, for example 

\begin{equation}
\vev{(x^2+y^2)g}_\rho=\frac{e^{-\frac{(x_1-x_2)^2}{2 \alpha ^2}}
   \left(\pi  \alpha ^4
   \left(2 \alpha^2+(x_1+x_2)^2\right)-4 \pi  \theta ^2
   \left((x_1-x_2)^2-2 \alpha^2\right)\right)}{8 \alpha^2}
   \end{equation}
   which is not positive definite.

In other words we see again the impossibility of operationalizing position measurements below the noncommutativity scale: even though the `particle' is supposedly five times the resolution of the device away, it overlaps with the device, though \emph{no} probability can be assigned to its detection. From the manifold point of view there is no appreciable pointwise overlap of the functions, so this effect appears to be non-local. Better, we should (again) recognize that we have a failure to operationalize position measurements below the noncommutatitivty scale, and that hence there really are no smaller regions, and no points, once the geometry is noncommutative. 

Working in the Moyal star (or `Weyl') representation greatly facilitates extracting the physical consequences of the theory because the usual methods of the calculus (and hence of standard field theory, including QFT) can then be applied. One `simply' has to multiply physical quantities, not in the usual way, but with the $\star$-product: an area is $x\star y$ (or the unequal $y\star x$) not $x\cdot y$; fields given as series expansions of the $x^i$s are to be understood in terms of expansions using the $\star$-product (for instance, exponentials); and terms in the equations of motion involve $\star$-multiplication. In other words, for every physical purpose, the $\star$-product is the relevant operation, and ordinary multiplication is only relevant insofar as it is involved in the definition (\ref{eq:moyal}) of the putatively `real' multiplication. In yet other words, the physical facts don't care about the commutativity of space, and it is thus natural to see it just as a convenient way of representing, the real noncommutative nature of space.

So it would then be a mistake to think, because there is such a representation, in terms of smooth functions on a manifold, that after all noncommutative geometry is a theory of perfectly ordinary space. As we have been at pains to show, the notions of arbitrary small regions and of points are not well-defined in any sense in such a theory. In the Weyl \emph{representation} of the algebra, the manifold is exactly that, a component of the representation, with excess representational structure for the true, essentially algebraic, fundamental objects. 

It's important to stress at this point that although elements of the algebra are represented by \emph{functions}, $\phi(x)$ over space, neither the value at any point, nor the restriction of $\phi(x)$ to any region corresponds to anything in the algebraic formulation, and so has any fundamental physical significance. (Let that sink in.) Normally we think of the value of a field at a given point as conveying some physical meaning, such as the electric field strength, but in the Weyl representation, this is not the case: only the full function corresponds to an element of the algebra and corresponds to something fundamental. Obviously this situation presents a puzzle, for most familiar physical quantities are associated with points and finite regions. We turn to this puzzle next.

\subsection{Finding Spacetime}\label{finding}
\label{sec:NCEC}

The Weyl representation provides a way of extracting empirical consequences from NCG. First, one can take the equations of motion for a classical field theory in ordinary commuting space, and rewrite them replacing all ordinary multiplication with Moyal star multiplication: the result is the equations of motion for a NCFT expressed in its Weyl representation. One obtains a more complex, but otherwise formally standard field theory -- though of course noncommutation of the coordinates undermines a straight-forward interpretation. At this point, we shift to an unequivocally quantum theory by formulating a `second quantized' NCFT. In particular, the machinery of the path integral formalism will be brought to bear so that standard methods allow the derivation of empirical results, especially probabilities for particles to scatter off one another in different ways -- scattering `cross-sections'.

These are often tacitly taken by physicists (at least those working in the QFT program) as basic concrete---indeed empirical---spacetime objects. That is, scattering occurs at an (extended) location so gives meaning to place; and depends, amongst other things, on the metric at that location, so gives meaning to geometry. Philosophically speaking, cross-sections are, crudely speaking, material objects to which a relationist might attempt to reduce space; or perhaps better, from which one might give a `dynamical interpretation' of spacetime. We don't take any stance on whether physicists have such programs in mind, but rather draw on the common tacit assumption in, for instance, string theory that our experiences of space can be recovered through scattering cross-sections, from a theory that is fundamentally not spatial. Again, the idea being that scattering gives empirical content to extended location and geometry; so recover a set of scattering amplitudes, and you have recovered spacetime. This construction is discussed in more detail in (\cite[chapter 7]{Nic:20}).

In particular, consider a world in which the cross-sections of a NCFT turn out to be correct. To understand the place of observed, classical space in such a NCFT it therefore makes good sense to focus on these cross-sections; there are other empirical aspects, and other ways in which space can be found, but cross-sections exemplify both perfectly. So we will proceed (and indeed speak) as if to understand the meaning of cross-sections in NCFT \emph{is} to understand classical spacetime. More could be said, but we expect it to be more of the same, and not to add to the central issue. And of course, we need to a way to understand `pointy' classical spacetime because we have argued that points and small regions have no physical meaning; and especially if one adopts an ontology in which the algebra is fundamental, for as we noted then neither points nor regions have fundamental significance. Our proposal here, then, is that recovering empirically meaningful cross-sections is to give derivative, empirical meaning to ordinary spacetime: to show how it `emerges'. (Though without reifying points.) This question has been addressed by Chaichian, Demichev and Presnajder (hereafter CDP) in a very interesting paper (\cite{chaichian2000quantum}, what follows is based on their analysis, though it suggests a somewhat different solution. 

The problem of finding scattering cross-sections can be further reduced to the calculation of `2-point functions': squared, these represent the probability that, left to itself, a quantum at $x$ in space and time would be `found' at $y$, the simplest kind of `scattering'.\footnote{Saying what `found' means in such contexts is to propose a solution to the measurement problem, something we are trying to sidestep as far as possible.} These, along with interaction terms, are the ingredients of the Feynman method for calculating cross-sections, so they can be taken as giving the empirical spatial content of a QFT---and hence of NCFT. The 2-point functions make the problem of giving a spatial interpretation very clear, for they are functions of $x$ and $y$, coordinates in phenomenal, commuting spacetime -- and so have no immediate significance in NCFT, in which the coordinates cannot be ordinary number-valued, since they don't commute! That is, finally, the question of the meaning of phenomenal cross-sections---so of \emph{space}---in NCFT narrows to the question of the significance of the \emph{commuting} coordinate arguments of the 2-point functions. 

Let us take a closer look at the 2-point functions: these are the vacuum state expectation values of a product of field operators---equation~\eqref{eq:moyal}. As we have emphasized, to this point we have not quantized the fields (in the narrow sense) by imposing a noncommutative algebra. Now we do take that step: take the state of the unquantized field, $\phi_W(x)$, which takes as its values field configurations in the Weyl representation; and second quantize it, promoting it to a quantum operator $\hat{\phi}_W(x)$. The corresponding 2-point function, $G_W(x_1,x_2)$, can be written
\begin{equation}
\label{ }
G_W(x_1,x_2) = \langle 0|\hat{\phi}_W(x_1)\hat{\phi}_W(x_2)|0\rangle.
\end{equation}
We quantize using the path integral formalism, in which such quantities are given by field integrals over the classical fields, weighted by the action:
\begin{equation}
\label{eq:wgf}
= \int D\phi_W \  \phi_W(x_1)\phi_W(x_2)\cdot e^{i\hbar\int d^nx\mathcal{L}(\phi_W(x),\dot{\phi}_W(x))}.
\end{equation}
Note that because we have fully quantized, $\hbar$ has finally made an appearance. 

This expression makes the interpretational issue very clear, for the dependence on the field at $x_1$ and $x_2$ is explicit in the (functional) integral. But, as discussed at the end of \S\ref{moweyl}, the point-value of a field in the Weyl representation has no fundamental meaning -- only the field configuration over the whole space represents anything in the algebraic formulation, namely an element of the algebra. So the same is true of the 2-point function: it can have some significance as a function over $\mathbb{R}^d\times\mathbb{R}^d$, but its point-values, or its integral over a region, do not. But these are exactly what we would like to take as scattering amplitudes, the empirical content of the theory.

A first response would be to more-or-less ignore this situation. One simply takes the coordinates in the Weyl representation to correspond to `phenomenal coordinates'---the ones by which we label points of ordinary, observed space.  At first glance it looks as if this response simply throws away the noncommutative spacetime and views the theory as one with unusual equations of motion; if the coordinates are just those of ordinary commuting space, then we just have a QFT in that space with a standard lagrangian modified by use of the Moyal star. However, while this approach might be expected to produce decent predictions over distances above greater than $\sqrt{\theta}$, it is conceptually incoherent in virtue of being undefined at distances less than $\sqrt{\theta}$, as we showed in \S \ref{spec}--\ref{operation}. Once again, the `pointlessness' of noncommutative geometry is the main point of this paper!

A second response, which recognizes this situation, to recover the appearance of commuting spacetime, is that based on the work of CDP \cite{chaichian2000quantum}. To keep things simple, we work in two dimensions, with $[\hat{x}_1,\hat{x}_2]=\theta$. As we saw, because of this noncommutivity, the Weyl field operators do not have the usual interpretation as localized quantities, but that doesn't mean that the same is true for other operators in the theory. Indeed, we should expect that some other observables do represent empirical spacetime quantities. 

In particular, we will consider Weyl fields that are `smeared' over a region of order $\theta$; the idea being that these are insensitive to sub-$\theta$ physics, while capturing the physics of super-$\theta$ regions. More specifically, and in the spirit of \S\ref{operation}, we propose that the following maps Weyl fields into the commuting fields of familiar empirical spacetime physics: i.e., fields describing observed scattering phenomena.\footnote{Technically, these fields are the `normal' symbols of the noncommuting fields.} 
\begin{equation}
\label{eq:pfield}
\phi_P(x) \equiv \int \mathrm{d}^2x^\prime\frac{e^{-(x-x^\prime)^2/\theta}}{\pi\theta}\phi_W(x^\prime).
\end{equation}
Smearing fields is a common practice in QFT, usually done to avoid pathological behaviours at short distances. Here we elevate it to a way to operationally use an object, a point, as an approximate avatar to connect with classicality. As can be seen from (\ref{eq:pfield}), the effect is to take a function of two variables, $x$ and $x'$ (in this case coordinates on two different spaces), and by integrating one out, return a function of just one.

This proposal is a `guess', a hypothetical part of the theory, subject to testing, and potentially to replacement by some other ansatz; but it is based on the most natural way of relating noncommuting and commuting space. (Note too that the form of the smearing is the simplest, rotationally invariant form one can have.) But we can take the $x$-arguments of these `new' fields to be those of observed space; while $x^{\prime}$ is the coordinate of the space in which the Weyl transforms live. That is,~(\ref{eq:pfield}) can be read as a map into the reals, from observed, commuting  space and Weyl field configurations (the integral means that the map depends on the full configuration): $x\times\phi_W(x^\prime)\to\mathbb{R}$.

Given (\ref{eq:pfield}) and the interpretation of $\phi_P(x)$, the 2-point function for the phenomenal fields is given by the path integral prescription:
\begin{eqnarray}
G_P(x_1,x_2) & \equiv & \langle 0|\hat{\phi}_P(x_1)\hat{\phi}_P(x_2)|0\rangle\\
 & = & \int D\phi_W \  \phi_P(x_1)\phi_P(x_2)\cdot e^{i\int d^nx\mathcal{L}(\phi_W(x),\dot{\phi}_W(x))}
 \end{eqnarray}
which is simply the smeared version of the Weyl 2-point function:

\begin{eqnarray}
 & = & \int\mathrm{d}^2x_1^\prime\mathrm{d}^2x_2^\prime \frac{e^{-(x-x_1^\prime)^2/\theta}}{\pi\theta}\cdot\frac{e^{-(x-x_2^\prime)^2/\theta}}{\pi\theta}G_P(x_1^\prime,x_2^\prime).\end{eqnarray}

Note that at this point we diverge from the CDP proposal. Their idea is that the action in the path integral should be rewritten in terms of the phenomenal field $\phi_P$. Their approach amounts to treating the phenomenal field as the true degrees of freedom. Instead, what we suggest is that we treat the Weyl fields as the true degrees of freedom, as we should if we take the noncommuting spacetime seriously: we simply recognize that the canonical degrees of freedom are not those we experience as phenomenal fields -- those are represented by $\phi_P$. Again, that hypothesis (in conjunction with the rest of the theory) is testable, and links the fundamental theory to experiment.

A puzzle arises, for if $\phi_P(x)$ has physical---if derivative---significance, then it seems as if its point values, and values over sub-$\theta$ regions do too; contrary to everything we have argued! But of course it is part of our proposal that they do not; only the differences over super-$\theta$ regions have physical meaning. The point values of $\phi_P(x)$ can only be understood as representational baggage, required by the formalism that we have adopted to formulate NCFT, and connect it to experience. (Of course, smearing enforces this interpretation, since it assumes the irrelevance of sub-$\theta$ physics.) Recall that the puzzle we are trying to solve is how to assign physical significance to \emph{any} regions if the fundamental objects are essentially algebraic, and non-spatiotemporal. The CDP ansatz addresses that issue, without giving significance to points. 

One could complain that the \emph{theoretical} meaning of the phenomenal field is unclear -- the $\phi_P(x)$ can formally be defined according to (\ref{eq:pfield}), but can we get a clearer insight into their place in the theory? In particular, do the phenomenal $x$s have an interpretation in the theory, since they are not the noncommuting coordinates? Since they label points of phenomenal space, an answer will illuminate how phenomenal space is found in a NCFT. CDP suggest an answer  (\cite{chaichian2000quantum},): they note that the phenomenal fields are equal to the expectation values of the noncommuting fields in so-called `coherent' states,~$|\xi_x\rangle$\footnote{Specifically, $|\xi_x\rangle=\exp{(x_1+ix_2})a^\dagger|0\rangle$}:
\begin{equation}
\label{eq:coh}
\phi_P(x) = \frac{\langle \xi_x|\hat{\phi}_W|\xi_x\rangle}{\langle \xi_x|\xi_x\rangle}
\end{equation}
Coherent states are the quantum states with a semi-classical behaviour. Here they take the added role of giving an operational meaning to the points of spacetime, albeit in their smeared guise. In the Weyl representation, a coherent state can be thought of as an isotropic state, centered on a point, $x$; the $x$s can be taken as their quantum numbers. Then (\ref{eq:coh}) shows how the point-values of the phenomenal field can also be understood as labelled by coherent states, taking the point in phenomenal space to be the corresponding quantum number.\footnote{CDP approach things from the other direction. They propose that the phenomenal fields should be labelled by the coherent states, and then conclude that they should be smeared according to (\ref{eq:pfield}).} 

But we will end the discussion here, having suggested how algebraic interpretations might be given, and shown how to recover the appearance of commutative spacetime, but with fields with no meaning below a certain scale---so again, pointless physics. 

\subsection{Empirical coherence and physical salience}\label{empiricalcoh}

More generally, Maudlin\cite{maudlin2007metaphysics} questions the feasibility of `deriving' classical spacetime from some non-spacetime theory (he has in mind deriving 3-space from $3N$-configuration space, but the point generalizes). At the heart of his concern is that even if a formal derivation can be found, involving a mathematical correspondence between classical spacetime structures and structures defined in terms of a (more) fundamental non-spatiotemporal theory, it does not follow that the classical spacetime just \emph{is} the more fundamental structure. Mathematical correspondences are too cheap: for instance, many systems are described by simple harmonic oscillator equations, but it would be a mistake to conclude that they were physically indistinguishable just because of this formal correspondence. According to Maudlin, for a reductive account, a formal derivation must also be `physically salient'. We take this to mean that the derivation must veridically track the way in which  fundamental structures `combine' to physically constitute  derivative ones. For instance,  in ideal gas theory  the  \emph{formal definitions} of `temperature' as mean kinetic energy and `pressure' as momentum transfer \emph{track} the corresponding phenomenal thermodynamical quantities: kinetic energy is transferred between the molecules of the gas to  liquid in a thermometer causing its expansion; and the pressure on the side of a vessel is due to the molecules contained colliding with it. The problem with a fundamental theory without spacetime is that our notions of what kinds of derivation track in this way are spatiotemporal notions, relying on colocation and dynamical interaction (of gas molecules with thermometers or vessel walls, say), for instance. But such familiar notions cannot apply if the physics involved is not, by supposition, itself spatiotemporal. So we face two problems: first, what new notions might apply? And second, even if we have a proposal, on what grounds can we conclude that we are correct?

If the analysis of this paper is correct, then noncommutative geometry is a nice example of this situation: the fundamental structure is algebraic, not a commutative geometry, and so concepts like `spatial location' are not primitives of the theory. Rather, spatial structure is derived. In particular, we have discussed the proposal that it be recovered via CDP ansatz, which we have also argued is not entailed by the theory, but an additional postulate. More precisely, it is an interpretational postulate, specifying how algebraic objects can `combine' to physically constitute classical spatial structures -- \emph{a  novel proposed non-spatiotemporal conception of which derivations are physically salient in the theory}. Thus the first problem can has been addressed in this case. (As in most cases, there is enough spatiotemporal structure in the underlying theory -- which is after all a deformation of a commutative geometry -- to find clues about how to reconstruct spacetime.)

As for the second problem, \cite{huggett2013emergent} proposes that such postulates, concerning how more fundamental structures compose to constitute less fundamental ones, are justified a posteriori, not a priori. (The paper briefly discusses NCFT and with other examples of theories without spacetime, along the lines found in this paper: identifying what spatiotemporal features are lost in each case, and explaining how they may be derived.) That is, how the fundamental gives rise to the less fundamental is not a matter of metaphysical necessity, but of physical contingency, and so is something that can only be discovered empirically, with the theory itself. For instance, if a theory of noncommutative geometry was empirically successful (in the usual ways, especially in making novel predictions that cannot be accounted for in any other known way) then both the theory \emph{and} CDP interpretational ansatz would be confirmed. That is, ultimately why we are justified in believing a derivation to be physically salient in the same way that any other scientific belief is justified: through successful confrontation with the data. No more is possible, but then it never is. 

Thus, in addition to introducing NCFT and raising some specific interpretational questions, this paper presents it specifically as an example of derived or `emergent' space, in order to illustrate and address Maudlin's challenge. There is a gap between noncommutative and commutative geometries, which can be \emph{formally} filled by the CDP ansatz; but if this strategy were empirically successful then we would have scientific grounds to further believe that the derivation is physically salient, that the ansatz is a veridical statement of physical composition. The hope is that working through the example makes that claim plausible, or at least intelligible.\footnote{Of course it is logically possible to deny it, but we would say (a) that the historical record contains examples of similar changes in the concept of physical salience, and (b) that such a denial would collapse into a general antirealism, which is not our target here -- rather the issue is whether there are special reasons to dispute the physical salience of derivations of spacetime.}

\section{Conclusion}\label{conc}

This paper had three goals beyond introducing noncommutative geometry to an audience of philosophers. The first was to convince the reader that one of the physical upshots of a theory having a noncommutative geometric structure is that it cannot include, in its ontology, a set of arbitrarily small spatial regions (\emph{a fortiori} points). We demonstrated this in two ways: in terms of the undefinability of arbitrarily small distances, using Connes' spectral triples, and in terms of the non-operationalisability of arbitrarily small separations, using measuring devices modelled by Gaussians in space. The second was to propose an appropriate ontology for a field theory set in a noncommutative space. We suggested that a field-first ontology was the only sensible option, and explored some of the consequences of this proposal. We were then led by this commitment to a puzzle about how to account for the appearance of a commutative spatiotemporal geometric structure, at least at the level of experience and experimental data. The final goal, then, was to discuss and assess a dynamical proposal, originally due to Chaichian, Demichev and Presnajder~\cite{chaichian2000quantum} for the emergence of spacetime using the resources of our best theory of matter--quantum field theory. 

As promised in the introduction, we demonstrated, via considerations of noncommutative geometries, one way of embedding a class of extant, well-confirmed physical theories (in this case, quantum field theories) in a broader logical landscape. This is, of course, a small step in the direction of unearthing all the important tacit commitments associated with interpretations of such theories. Even so, we have made some real progress on that front. We realised that we can, in fact, talk about spacetime without it being indiscrete.

\subsubsection*{Acknowledgements}
We would like to thank Jeremy Butterfield, Neil Dewar, Gaetano Fiore, James Read, Tiziana Vistarini, the audience at the QISS seminar, and two anonymous reviewers for very helpful comments and feedback. FL acknowledges the support of  the INFN Iniziativa Specifica GeoSymQFT and the Spanish MINECO under project MDM-2014-0369 of ICCUB (Unidad de Excelencia `Maria de Maeztu'). This publication was made possible through the support of grants from the John Templeton Foundation. The opinions expressed are this of the authors, and do not necessarily reflect the views of the John Templeton Foundation. The work was also supported by a Collaborative Fellowship from the ACLS, and a fellowship from the Institute of Philosophy at the University of London.

\end{document}